\documentstyle[aps,pre,epsfig]{revtex}
\begin{document}

\draft

\title{Two-point correlation functions of the diffusion-limited 
annihilation in one dimension}

\author{Su-Chan Park,${}^1$ Jeong-Man Park,${}^2$ and Doochul Kim${}^1$}

\address{${}^1$ School of Physics, Seoul National University, Seoul 151-747,
Korea}

\address{${}^2$ Department of Physics, The Catholic University of
 Korea, Puchon 420-743, Korea}

\date{\today}
\maketitle
\begin{abstract}
Two-point density-density correlation functions for the
diffusive binary reaction system $A+A\rightarrow\emptyset$ are
obtained in one dimension via Monte Carlo simulation.
The long-time behavior of these correlation functions
clearly deviates from that of a recent analytical
prediction of Bares and Mobilia [Phys. Rev. Lett. {\bf 83}, 5214 (1999)].
An alternative expression for the asymptotic behavior is 
conjectured from numerical data.
\end{abstract}


\section{Introduction}
Nonequilibrium systems have been studied extensively during the past
decades. Reaction-diffusion systems especially have attracted much
interest due to their relevance to many areas of physics, biology, economics,
and so on\cite{Pri97,MD99}. 
A commonly used description of the dynamics of reaction-diffusion
systems is the master equation.
The master equation for the probability distribution of a many-body
system is a linear equation, and can be written as an imaginary time
Schr\"odinger equation with a (generally non-Hermitian) 
Hamiltonian acting on a many-particle Fock space\cite{KS68}. 
The energy spectrum of the Hamiltonian contains all the information
about the system.
It is, in general, difficult to diagonalize the Hamiltonian, and
many methods to treat the systems 
have been developed: the mean field rate equation,
the scaling arguments\cite{KR85}, and the renormalization group (RG)
technique\cite{Lee94,PKP00} to name only a few.

In simple situations, it is possible to find exact solutions
of the model systems. 
A prototype of such models is the one-dimensional diffusion-limited
pair annihilation and/or creation of hard core particles.
The model can be solved exactly only
in the so-called free-fermion limit\cite{Lush87,GS95,Sch96}.
As implied in the work of Spouge\cite{Spo88}, the free
fermion limit, for example, of the pair annihilation model
can be reduced to a one-dimensional
diffusion problem of a noninteracting ``free'' particle.
Any exact results for models beyond the free-fermion case would add much 
to our understanding
of the reaction-diffusion systems of hard core particles.

In this context, a recent work by Bares and Mobilia\cite{BM99}
is of interest.
Studying the model mentioned above in the
general parameter space, Bares and Mobilia suggested a method
to find an exact solution of a general one-dimensional reaction-diffusion
system applicable to the non-free-fermion case.
The method relies on an analog of the Wick theorem.
The present authors, however, pointed out that
the Wick theorem does not hold except for the free-fermion case\cite{PPK00}.
According to Ref.\cite{PPK00}, the method of Bares and Mobilia
relies on a neglect of the higher order correlations, so 
the result of Ref.\cite{BM99} can at best be regarded as a Hartree-Fock-type
approximation.

What is unknown is the validity of the approximation. In particular, 
the long-time behavior of the density and the correlation reported
in Ref.\cite{BM99} could be correct in merely a 
fortuitous way. To check this possibility, we performed an extensive
Monte Carlo simulation on the system (with no bias) and compared 
the numerical results with the analytical predictions of Ref.\cite{BM99}.
We find that the numerical result and the analytical prediction
of Ref.\cite{BM99} for the density are consistent with each other 
in the long-time limit, 
while those for the two-point density-density correlation
functions clearly deviate from one another. We conjecture from the
data that the two-point density-density correlation 
function $M_r(t)$, defined below, contains an extra term with a simple
analytic structure, and we estimate its coefficient.
We also provide a heuristic argument as to why the density behavior
is not modified in the Hartree-Fock-type approximation.

\section{model}
In this Brief Report, we study the standard reaction-diffusion system
$A+A\rightarrow \emptyset$ with symmetric hopping on a one-dimensional 
lattice of size $L$ with periodic boundary conditions. Each site
is either occupied by a hard core particle or is vacant.
We restrict ourselves to the dynamic rules
\begin{equation}
\begin{array}{ccc}
A\emptyset \rightarrow \emptyset A & {\rm with~rate}& {1\over2},\\
\emptyset A \rightarrow A \emptyset & {\rm with~rate}&{1\over2},\\
AA \rightarrow \emptyset\emptyset&{\rm with~rate}&p,
\end{array}
\end{equation}
where, in what follows, $p$ is limited to be smaller than or equal to $1$. 
To implement the above dynamics on simulation, we use the following algorithm:
At time $t$, we choose one of the occupied sites randomly,
and select one of the
nearest neighbors of the chosen site with equal probability $1 / 2$.
If the selected nearest neighbor is empty, the chosen particle
hops to that site. Otherwise, the  two particles are annihilated with 
a probability $p$, and with a probability $1-p$ nothing happens.
After an attempt, the time is increased by $1/N(t)$, where
$N(t)$ is the total particle number at a time $t$.
The corresponding master equation of this algorithm 
can be written as ${\partial_t}|\Psi;t
\rangle = - H |\Psi;t\rangle$, with the Hamiltonian
\begin{eqnarray}
H = -\sum_{m=1}^{L} \Biggl [ {1\over 2} \sigma_{m+1}^+ \sigma_{m}^-
+{1\over 2}\sigma_{m+1}^- \sigma_{m}^+ + p \sigma_{m}^- \sigma_{m+1}^- 
-\sigma_{m}^+ \sigma_{m}^-  - \gamma \sigma_{m}^+ \sigma_{m}^-
\sigma_{m+1}^+ \sigma_{m+1}^- \Biggr],
\label{hamiltonian}
\end{eqnarray}
where $\gamma \equiv p - 1$ and $\sigma_m^\pm$ represent the 
Pauli matrices. 
We regard the spin-down (-up) state as a vacuum (particle) state.
$| \Psi;t\rangle$ is defined as
$|\Psi;t \rangle \equiv \sum_{\eta} P(\eta,t) |\eta\rangle$, where $\eta$
is the representation of a possible microscopic configuration, and $P(\eta,t)$
is the probability with which the system is in state $\eta$ at
time $t$\cite{KS68}.
The free-fermion limit corresponds to the $\gamma=0$ (or $p=1$) case.
The above Hamiltonian, with a fully occupied 
initial condition, was studied in Ref. \cite{BM99} and 
it was reported that the asymptotic behavior
of density and correlation functions up to the subleading order become
(we are using our notation)
\begin{eqnarray}
\rho^{\rm BM}(t) &\simeq& {1 \over \sqrt{4 \pi t}} + { 1 - p\over \pi p t},
\label{denanti}\\
{\cal C}^{\rm BM}_r(t) &\simeq& -{1 \over 4 \pi t} + 
\left \{ \pi r + 8 { p - 1 \over p } \right \}
{1 \over (4 \pi t)^{3/2}},
\label{connected}
\end{eqnarray}
respectively, where $\rho(t)$ is the density at time $t$, ${\cal C}_r(t)$ 
is the 
two-point (connected) correlation function, and the superscript
BM stands for ``the result of Bares and Mobilia''.

In the following, we concentrate our attention on the two-point density-density
correlation function, which is defined as
\begin{equation}
M_r(t) = {1 \over L} \sum_m \langle n_m n_{m+r} \rangle(t) = {\cal C}_r(t)
+ \rho(t)^2,
\end{equation}
where $n_m \equiv  \sigma_m^+ \sigma_m^- $, and $\langle
\cdots \rangle(t) $ represents the ensemble average at time $t$.
The reason for studying $M_r(t)$ rather than $C_r(t)$ is twofold: First, 
the subleading order of the connected correlation function is
the same as the leading order of $M_r(t)$ [see Eqs. (\ref{conjmom}) and 
(\ref{conjcon})]
and, in general, the study of the leading order is more reliable than that of 
a subleading order.
Second, $M_1(t)$ is an order parameter of the 
pair contact process with diffusion (PCPD) model\cite{pcpd}, and it is
known\cite{HS00} that the PCPD model in the absorbing phase shares features with
the annihilation model studied here. 
From Eqs. (\ref{denanti}) and (\ref{connected}), the leading behavior of
$M_r(t)$ is meant to be
\begin{equation}
M_r^{\rm BM}(t)   \simeq {\pi r \over (4 \pi t )^{3/2}}.
\label{anti_mom}
\end{equation}

To see how good Eqs. (\ref{denanti}) and (\ref{anti_mom}) are,
we present the numerical results in Sec III.

\section{numerical results}
We performed the Monte Carlo simulations for several $p$'s
($p=1$, $3/4$, $1/2$, and $1/4$) with a system size
$L=10~000$.
The initial conditions for all cases are fully occupied states
as in Ref.\cite{BM99}. 
For each $p$, we realize $10^6$ samples. 
We monitored the simulation up to  $10^5$ Monte Carlo time steps. 

At first,
we compare the density obtained by numerics 
with Eq. (\ref{denanti}) in Fig. \ref{density}. 
This figure clearly shows that 
Eq. (\ref{denanti}) is in excellent agreement with the numerical results
up to subleading orders. This result also implies that a system size of
$10~000$ is large 
enough not to show the finite size effect up to $10^5$ time steps.

Next we compare the numerical results of 
the density-density correlation functions 
with Eq. (\ref{anti_mom}). For this purpose, in Fig. \ref{moment} we draw 
$M_r(t)/M_r^{\rm BM}(t)$  
as a function of time rather than $M_r(t)$ itself.
If Eq. (\ref{anti_mom}) is correct, all data are expected to 
converge to the constant value of 1 in the long-time limit,
but no such convergence
is observed except for $p=1$, i.e., the free-fermion case and for
large $r$.
Figure \ref{moment} clearly shows deviations 
from that anticipated by Eq. (\ref{anti_mom}).
Hence the correct asymptotic behavior of the density-density
correlation function may take the form 
\begin{equation}
M_r(t) \simeq {\pi r \over (4 \pi t )^{3/2}} \Delta(r,p).
\end{equation}
One expects $\Delta(r,p=1) = 1$ for all $r$.
We estimate $\Delta(r,p)$ from the data by the least squares fits, and show them
in Fig. \ref{conjecture} as functions of $1/r$
for various $p$'s. It appears that $\Delta(r,p)$  has a very simple 
dependence on $1/r$:
\begin{equation}
\Delta(r,p) = {\lambda (p) \over r} + 1.
\end{equation}
Interestingly, $\lambda (p)$ also seems to have a simple mathematical
structure, i.e., $\lambda (p) \propto ( 1 - p)/p$;
see the inset of Fig. \ref{conjecture}. 
Combining the results, we conjecture the forms of $M_r(t)$ and ${\cal C}_r(t)$
to be
\begin{eqnarray}
M_r(t) &\simeq& {1 \over (4 \pi t)^{3/2}} \left (
\pi r + c { 1 - p \over p} \right ),
\label{conjmom}\\
{\cal C}_r(t) &\simeq& - {1 \over \sqrt{4 \pi t}} + \left \{
\pi r + (8 - c) { p - 1 \over p}\right \} { 1 \over (4 \pi t)^{3/2}}
\label{conjcon}
\end{eqnarray}
with $c = 3.4 \pm 0.2$ in case of the fully occupied initial condition. 

In the long-time limit and at large length scales, all the $p$ dependence is 
suppressed in the density and correlation functions. This observation
is evidence of the irrelevance of $\gamma$ in the RG sense.
In Sec. IV, we use this irrelevance argument to present
a possible explanation for
the partial success of Ref.\cite{BM99}.

\section{discussion and summary}
This section discusses the success and failure of the method given in
Ref.\cite{BM99}, and summarizes our work.
First of all, the incorrect result of $M_r(t)$ is ascribed to the non-Gaussian
form of the generating function (GF)\cite{PPK00}.
The relation between the Gaussian form of the GF and 
the Wick factorization is explicitly shown in Appendix \ref{gauss}.
Furthermore, Appendix \ref{FPAMex} explicitly shows that 
Wick factorization is not always possible, even though
the probability conservation is satisfied.
From this point of view, the method of Bares and Mobilia may
be regarded as an 
approximation scheme where quartic terms are neglected.
As long as only the decay exponent of $M_r(t)$ is concerned, Eq.
(\ref{anti_mom}) is good enough. 
The irrelevance of $\gamma$ in the RG sense makes it possible for one
to obtain the correct decay exponents of density and correlation functions
from the exactly solvable $\gamma=0$ limit.

We also resort to the irrelevance argument to explain the correct subleading
behavior of the density. First note that the quartic coupling in 
the generating function,
say $W$, contributes to $M_r$ for nonzero $\gamma$, but this contribution
should be  $O(t^{-3/2})$, 
because $W$ is generated by the 
irrelevant operator $\gamma$ and $M_r \sim O(t^{-3/2})$.
Thus we expect that the implicit contribution
of $W$ to the density is at most $O(t^{-3/2})$ which happens to be 
smaller than the subleading order of the density decay. 
The Wick factorization, which neglects $W$, then
gives a correct time dependence of the density 
up to the subleading order. We also believe that the method in Ref.\cite{BM99}
would be partially 
successful only if the nonquadratic terms in the Hamiltonian
are irrelevant in the RG sense.
 
For completeness, we address the question of
why the Wick factorization may yield 
the correct steady state density in the presence of pair creation. 
When the pair creation rate is not zero, 
the mean field rate equation for the density can be written as
\begin{equation}
{d \rho \over dt} = -2 \epsilon' \rho^2 + 2 \epsilon ( 1 - \rho )^2,
\end{equation}
with $\epsilon'$  the annihilation rate and $\epsilon$  the
pair creation rate, using the same notation as in Ref.\cite{BM99}.
Its stationary state solution is
$\rho_{\rm s} = 1 / ( 1 + \sqrt{\epsilon' / \epsilon})$, which interestingly
is the same as the exact solution\cite{dMdO98}. This result implies
that the exact 
steady-state solution displays the characteristics of the
mean field solution. Hence an
exact result for the steady-state density from the Wick 
factorization is not surprising. 

In summary, we have extensively 
simulated the diffusion-limited annihilation model,
and have  
shown numerically that the long-time behavior of the correlation
functions given in  Ref.\cite{BM99} is not
correct, although the behavior of density is.
In addition, an analytical form of
the asymptotic behavior of the correlation functions is conjectured.

\section*{Acknowledgments}

This research was supported by the Catholic University of Korea
research fund 2000, and by the Brain Korea 21 Project at Seoul
National University.

\appendix
\section{Generating function and Wick theorem}
\label{gauss}
This appendix proves that the Gaussian form of the generating function (GF)
is a sufficient and necessary condition for Wick factorization.
We restrict ourselves to the even sector of the Hilbert space as
in Ref.\cite{BM99}. In the following, we adopt the notation of 
Ref.\cite{PPK00}. First let us assume that the GF,
discussed in Refs.\cite{PPK00} and 
\cite{SSS96} takes the Gaussian form
\begin{equation}
Z[\xi] = \exp\left [\sum_{q_1< q_2} \xi_{q_1} \xi_{q_2} f(q_1, q_2) \right],
\end{equation}
with $\xi_q$ the Grassmann numbers.
Proving the Wick factorization for this GF is a simple
combinatoric problem. One can easily check that the Wick theorem
can be applied; hence we prove that the Gaussian form of the GF
yields Wick factorization. 

Next we consider the following question: Does the Wick factorization  
imply a Gaussian form of the GF?
To answer this, we adopt the  {\it  reductio  ad absurdum}.
Let us assume that   the GF does not take
a Gaussian form, but is instead of the form 
\begin{eqnarray}
Z[\xi] = \exp \Biggl [\sum_{q_1< q_2} \xi_{q_1} \xi_{q_2} f(q_1, q_2) +
 \sum_{q_1 < q_2 < q_3 < q_4} \xi_{q_1}
\xi_{q_2}\xi_{q_3}\xi_{q_4} W(q_1, q_2 , q_3, q_4) + \cdots \Biggr ],
\end{eqnarray}
where $W(q_1,q_2,q_3,q_4)$ is the quartic coupling, and $\cdots$ stands for
higher order terms.
One can calculate the four operator correlation functions simply
by differentiation, and find
\begin{eqnarray}
\langle a_{q_1} a_{q_2} a_{q_3} a_{q_4} \rangle =
\langle a_{q_1} a_{q_2} \rangle \langle  a_{q_3} a_{q_4} \rangle -
\langle a_{q_1}a_{q_3} \rangle\langle a_{q_2} a_{q_4} \rangle 
+
\langle a_{q_1} a_{q_4} \rangle  \langle a_{q_2} a_{q_3} \rangle
+ W(q_1, q_2 , q_3, q_4)
\end{eqnarray}
for $q_1<q_2<q_3<q_4$.
The presence of $W$ prohibits the factorization.
Since this result
stems from the assumption of a non-Gaussian form of the GF,
we proved that the Wick factorization implies a Gaussian
form of the GF.

Hence the Gaussian form of the GF is a
sufficient and necessary condition of the Wick factorization.

\section{Four-Particle Annihilation Model}
\label{FPAMex}
In this appendix,
we introduce a simple toy model, called the  
four-particle annihilation model (FPAM), defined in one 
space dimension, and analyze it to show that the probability conservation 
does not imply Wick factorization.
The dynamics occurs only when four particles form
a cluster. If a four-particle cluster exists, it
annihilates with a rate $\lambda$. Hopping events are not allowed.
The Hamiltonian of the FPAM is
\begin{eqnarray}
H_{\rm FPAM} = -\lambda \sum_{n} \bigl [ \sigma_n^- \sigma_{n+1}^-
\sigma_{n+2}^-\sigma_{n+3}^-
- \sigma_n^+ \sigma_n^-
\sigma_{n+1}^+ \sigma_{n+1}^-\sigma_{n+2}^+ \sigma_{n+2}^-
\sigma_{n+3}^+ \sigma_{n+3}^- \bigr ],
\end{eqnarray}
 where the periodic boundary condition is assumed. 
For simplicity, let us consider a fully occupied initial condition with a system
size $4$,
that is, $|\Psi;0 \rangle = \prod_{i=1}^4 \sigma_i^+ |\emptyset \rangle$,
where $|\emptyset  \rangle$ is the particle vacuum state and $|\Psi;t \rangle
$ is the state vector as usual.
In the momentum space  $|\Psi;0 \rangle = \prod_{q>0}
a_q^\dag a_{-q}^\dag |\emptyset
\rangle $, where possible values of $q$'s 
are ${\pi /4}$ and ${ 3 \pi / 4}$.
The state vector at time $t$ is $|\Psi;t \rangle = e^{-4 \lambda t}
|\Psi;0 \rangle + (1 - e^{-4 \lambda t} ) |\emptyset\rangle$. 
The nonvanishing two-operator correlation functions are
$\langle a_{-{\pi / 4}} a_{\pi / 4} \rangle(t)$ and
$\langle a_{-{3 \pi / 4}} a_{3 \pi / 4} \rangle(t)$ with
the value $e^{-4 \lambda t}\tan {\pi / 8}$ and $e^{-4 \lambda t}
\cot { \pi / 8}$, respectively. The four-operator
correlation function at time $t$ is
$\langle a_{-{3 \pi / 4}} a_{3 \pi / 4}
a_{-{\pi / 4}} a_{\pi / 4} \rangle(t)=e^{-4 \lambda t}$ which
is different from $\langle a_{-{\pi / 4}} a_{\pi / 4} \rangle(t)
\langle a_{-{3 \pi / 4}} a_{3 \pi / 4} \rangle(t)$ by
$e^{-4 \lambda t} - e^{-8\lambda t}$: Hence, we see a  
failure of the Wick theorem under the condition of probability conservation.
In other words, probability conservation has nothing
to do  with Wick factorization.

\begin{figure}
\caption{Time dependence of the density in a log-log plot. 
The time is measured  in units of Monte Carlo steps (MCS), and
the lattice constant is set to 1.
The values of $p$ for 
the solid lines are
$1,0.75,0.5$, and $0.25$ from the bottom.
The broken line is the anticipated leading
behavior $1/\sqrt{4\pi t}$.
Inset: time dependence of $\rho(t) - 1/\sqrt{4\pi t}$ for various
$p$'s ($p=0.75, 0.5$, and $0.25$ from the bottom). The broken lines are 
$(1-p)/(\pi p t)$, and show excellent agreement with the
numerical results.}
\label{density}
\end{figure}

\begin{figure}
\caption{Plots of $t$ vs $M_r (t)/M_r^{\rm BM} (t)$ 
for various $p$'s. If 
Eq. (\ref{anti_mom}) is correct, all data must converge to
$1$ as $t$ goes to infinity. However, except for $p=1$ (that is $\gamma=0$), 
the data show a clear discrepancy. 
In this figure, we present the simulation
results for $r=2$ ($\Diamond$), $5$ ($\bigtriangleup$),
 $8$ ($\bigtriangledown$),
and $50$ ($\bigcirc$).}
\label{moment}
\end{figure}

\begin{figure}
\caption{Estimated values of $\Delta(r,p)$ as a function of $1/r$ 
for $p$ = 0.25 ($\bigcirc$), 0.5($\bullet$), 
0.75($\bigtriangleup$), and 1.0 ($\bigtriangledown$). 
Error bars stand for standard deviations of the least squares fits of Fig.
\ref{moment}.
As in Fig. \ref{density}, the lattice constant is set to 1.
Each data set for the same $p$
seems to lie on a straight line. 
The fitted lines are also shown.
Inset : we draw the slopes of the
fitted lines as a function of $p$. These slope values also fall on a 
straight line (solid line), whose slope is $1.08\pm0.03$.
}
\label{conjecture}
\end{figure}

\centerline{\epsfxsize =\textwidth \epsfbox{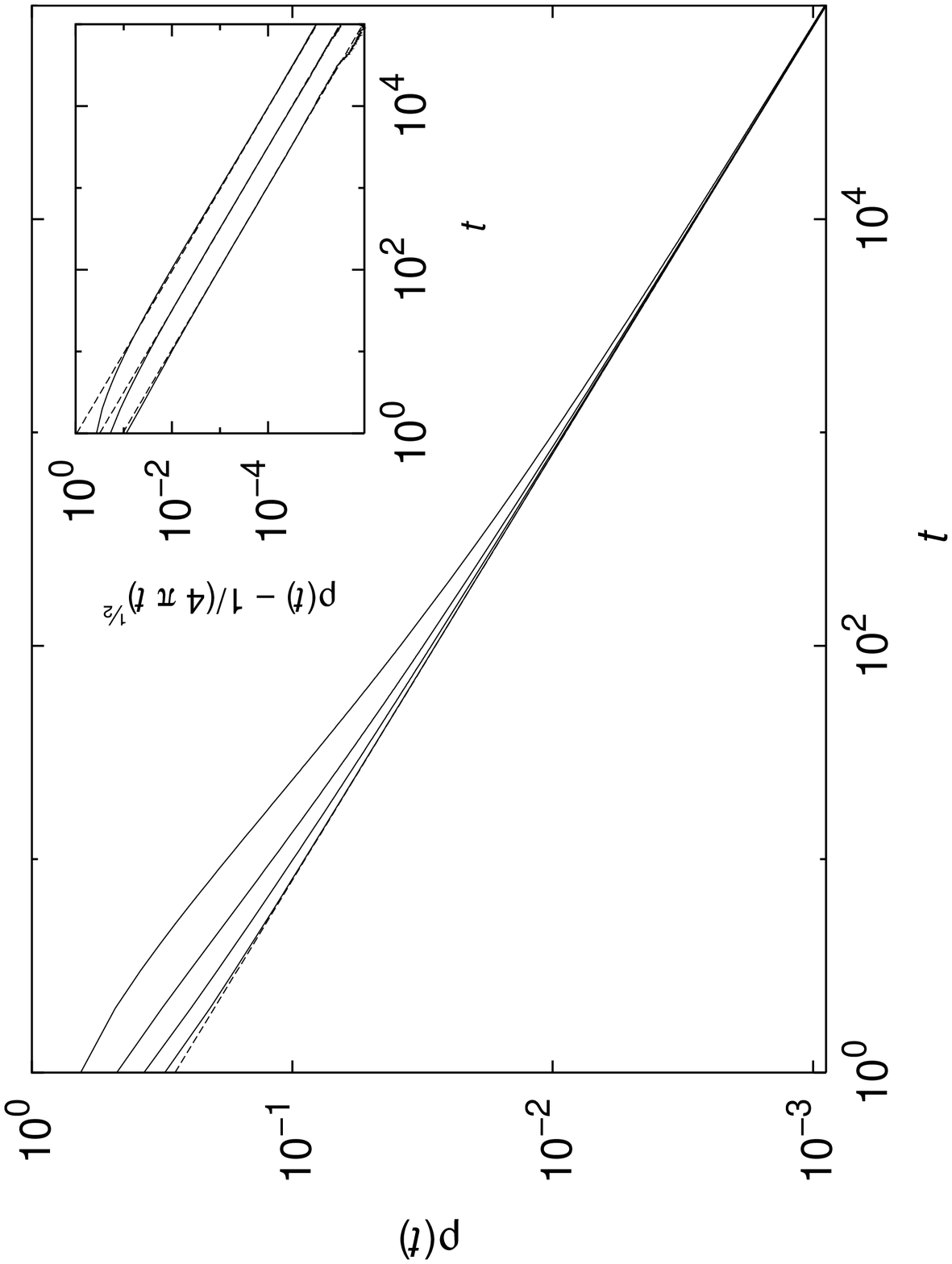}}
\centerline{\epsfxsize =\textwidth \epsfbox{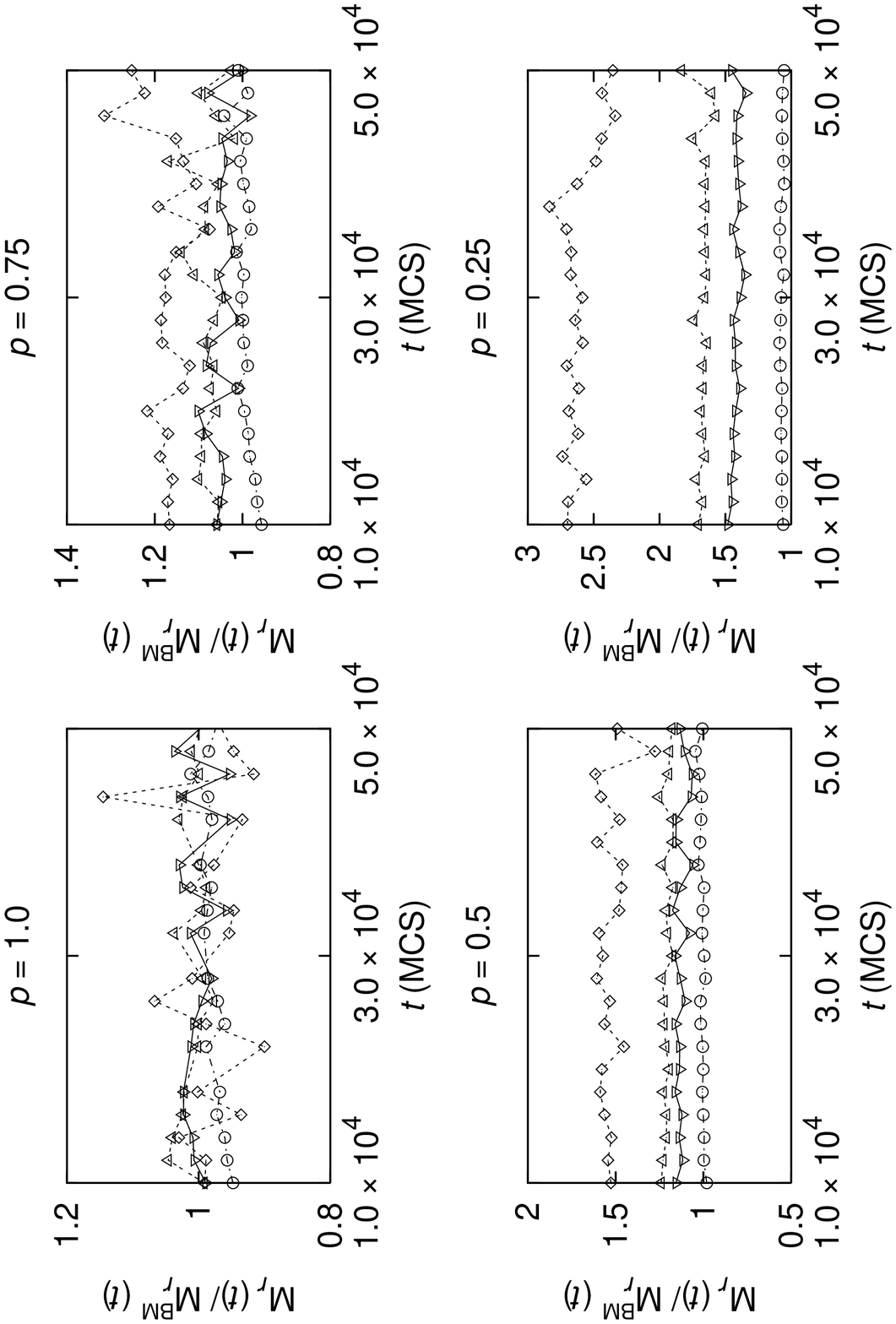}}
\centerline{\epsfxsize =\textwidth \epsfbox{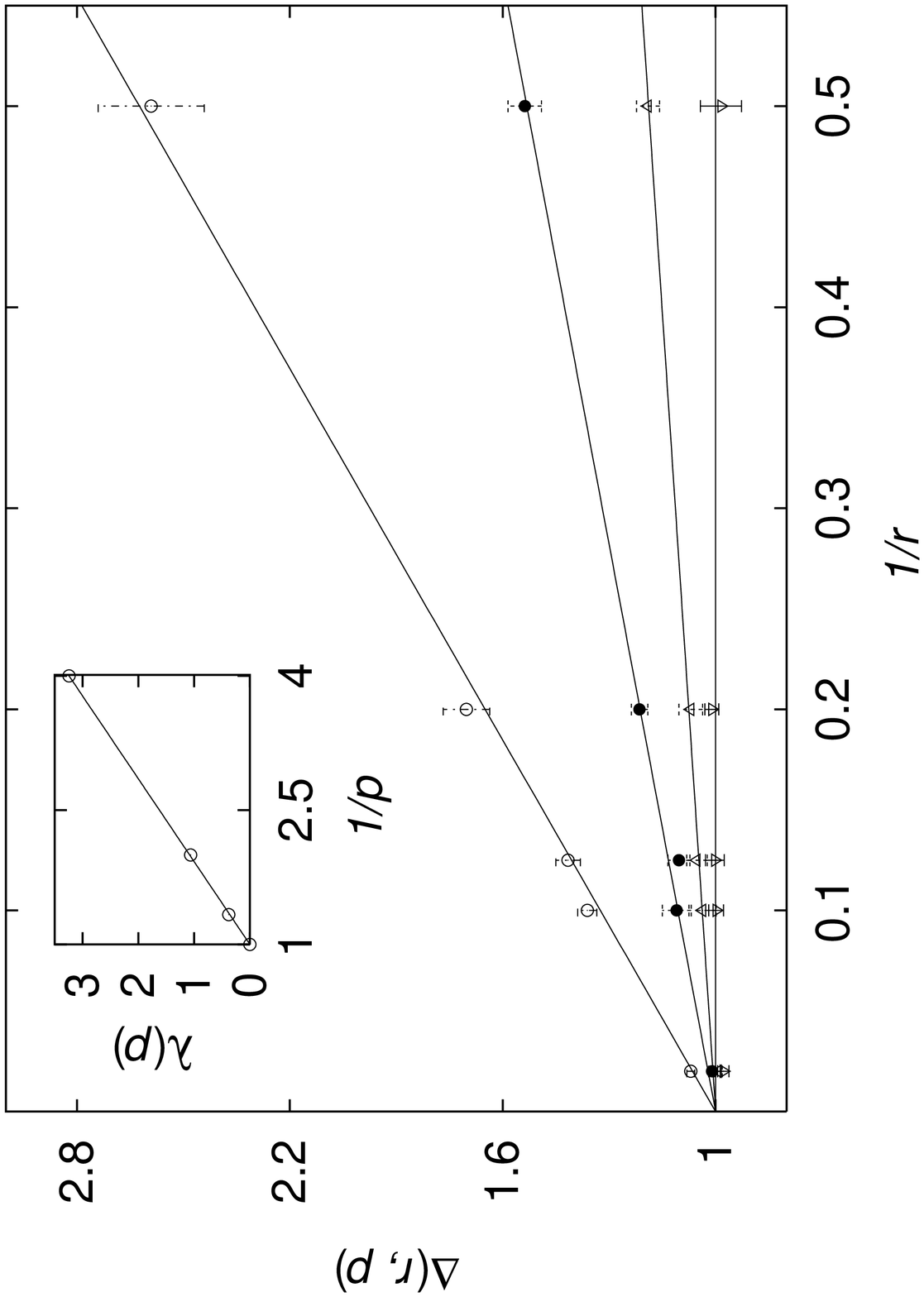}}
\end{document}